\documentstyle[epsfig]{aipproc}

\def\Wvis {$W_{\mathrm{vis}}$}
\newcommand{\GG}{\gamma ^* \gamma ^*}
\newcommand{\ee}{\rm{e}^{+} \rm{e}^{-} }
\newcommand{\gaga}{\gamma \gamma}
\newcommand{\qqbar}{q \bar{q}}
\newcommand{\ra}{\rightarrow}
\newcommand{\pz}{$\pi ^0 \,$}
\newcommand{\px}{\pi ^0 \, + \, X}
\newcommand{\ks}{$K_S^0 \,$}
\newcommand{\kx}{K_S^0 \, + \, X}
\newcommand{\Wgg}{$W_{\gaga}$}

\newcommand{\dspt} {$d\sigma / dp_T \,$}
\newcommand{\dseta} {$d\sigma / d|\eta| \,$}
\newcommand{\pt}{$p_T \,$}

\begin{document}
\title{Inclusive \pz and \ks \, in $\gaga$ reactions at L3}
\author{Pablo Achard $^*$ }
\address{University of Geneva, DPNC, 24 quai E. Ansermet, 1205 Geneva, Switzerland.\\
$^*$ on behalf of the L3 collaboration}

\maketitle

\begin{abstract}
The $\rm{e}^{+} \rm{e}^{-} \rightarrow \rm{e}^{+} \rm{e}^{-} \px$
and $\rm{e}^{+} \rm{e}^{-} \rightarrow \rm{e}^{+} \rm{e}^{-} \kx$
reactions are studied at LEP 
using data collected with the L3 detector 
at $\sqrt{s}=189-202  ~ \rm{GeV}$. Preliminary results for the  differential cross section $d\sigma /dp_T$ 
in the transverse momentum range $0.2 ~ \rm{GeV} < p_T < 7.5 ~ \rm{GeV} $ 
at  central values of   pseudo-rapidity, $|\eta| < 0.5$ or $|\eta| < 1.5$, 
and the differential  cross section $d\sigma / d|\eta| $ for different  
values of $p_T$ are presented.
For $ p_T \le 1 ~ \rm{GeV} $, the $p_T -$dependence of the data are well described by an exponential fit.
For $ p_T > 1 ~ \rm{GeV} $ the cross sections are
compared to NLO QCD calculations and to LO Monte Carlo predictions.
\end{abstract}

\section*{Introduction}

Two-photon collisions are the main source of hadrons at LEP 
 via the process $\rm{e}^{+} \rm{e}^{-} \rightarrow \rm{e}^{+} \rm{e}^{-}
 \gamma ^{*}  \gamma ^{*}  \rightarrow 
 \rm{e}^{+} \rm{e}^{-} $ {\sl hadrons} .
In the Vector Dominance Model (VDM), each photon can fluctuate
into a vector meson with the same quantum numbers as the photon, 
thus initiating a strong interaction process with characteristics 
similar to hadron$-$hadron interactions. This
process dominates in the "soft" region, where hadrons are produced
  with a low  transverse momentum \pt.
Hadrons with  high \pt {} are produced by the QED direct process, $\GG \ra \qqbar$, 
or by QCD processes originating from the hadronic content of the photon.
 Depending on wether one or two photons fluctuate into a hadronic system,
 the QCD processes are called single$-$resolved or double$-$resolved respectively.
In two-photon interactions,  
  QCD calculations are available 
 for single particle inclusive production at Next-to-leading order (NLO) \cite{kniehl}.
\par
In this paper the inclusive \pz production and the inclusive \ks \, production from
quasi$-$real photons are measured. The  
center-of-mass energy  of the two interacting photons, \Wgg , is 
greater than 5 GeV for \pz and greater than 10 GeV for \ks \@. The $\pi^0$ is measured 
in the transverse momentum range $0.2 \le p_T \le 7.5 ~ \rm{GeV}$ and in the pseudo$-$rapidity
interval
 $ -.5  \le \eta \le .5$, with $\eta  =  -\ln (\tan \theta / 2 )$. The angle
 $\theta$ is the polar angle of the \pz, relative to the beam axis. 
 The \ks is measured in the range $0.2 \le p_T \le 4 ~ \rm{GeV}$ and $ -1.5  \le \eta \le 1.5$
 
 The differential cross sections are compared to the Monte 
 Carlo models  PHOJET \cite{Engel} and  PYTHIA \cite{PYTHIA} and to  analytical 
 NLO QCD calculations \cite{kniehl}.

\section*{Event selection}
The data used for this analysis were collected by the L3 detector \cite{L3}
in 1998 and 1999 at beam energies from 189 GeV to 202 GeV, with an average 
energy value of 194 GeV.
The integrated luminosity is 414 ${\rm pb}^{-1}$.
\par
The   $\rm{e}^{+} \rm{e}^{-} \rightarrow 
 \rm{e}^{+} \rm{e}^{-} $ {\sl hadrons} 
 processes are simulated with the 
   PHOJET\footnote{PHOJET version1.05c}
 and PYTHIA\footnote{PYTHIA version 5.718 and JETSET version 7.408}  
event generators.
For the background 
 the annihilation  processes 
$\ee \rightarrow$ {\sl hadrons}($\gamma $),
ZZ($\gamma $), Zee($\gamma $), We$ \nu$($\gamma $)
are simulated  with  PYTHIA \cite{PYTHIA}; 
  KORALZ \cite{KORALZ} is used for $\ee \rightarrow \tau^{+} \tau^{-}(\gamma )$
and KORALW \cite{KORALW} for $\ee \rightarrow W^{+} W^{-}$.
For the  $\ee \ra {} \ee \tau^{+} \tau^{-}$ channel 
the generator  DIAG36 \cite{DIAG36} is used.
 The events 
are simulated in the L3 detector using the GEANT \cite{GEANT}
and GEISHA \cite{GEISHA} programs
and 
passed through the same reconstruction program as the data.

\par
The selection of  $\rm{e}^{+} \rm{e}^{-} \rightarrow 
 \rm{e}^{+} \rm{e}^{-} $ {\sl hadrons} events is based on information from 
the central tracking detectors and from 
 the electromagnetic (BGO)  and hadronic   calorimeters. The following cuts are applied:

\begin{enumerate}
\item To reject annihilation events, the total energy in the  
   calorimeters
is required to be less than 40 \% of the center-of-mass energy.

\item To exclude radiative events of the type
 $\ee {} \ra \gamma  Z \ra \gamma \qqbar$, the total energy 
in the electromagnetic   calorimeter must be lower than 50 GeV.

\item Hadronic events are selected with at least 6 particles in the detector.
These particles can be charged tracks  or 
isolated clusters in the   calorimeters.

\item Quasi$-$real two-photon interactions are selected by 
excluding  events  with a cluster of energy greater than 70 GeV
and polar angle greater than 33 mrad.

\item Beam-gas and beam-wall events are suppressed
by the requirement that at least 500 MeV must be deposited in the  electromagnetic   calorimeter.

\item The analysis is limited to events with  a visible hadronic mass \Wvis $> 5 ~ \rm{GeV}$ , to 
allow comparisons with the Monte Carlo generators which generate events with 
\Wgg $> 3 ~ \rm{GeV}$. \Wvis , is the effective mass of the hadronic
system, calculated from the four-momentum vector of all the measured particles\footnote
{All particles are considered to be pions, except for isolated clusters in the
electromagnetic calorimeter which are considered to be photons.}.

\end{enumerate}

After these cuts, the  background  ($ \sim $ 1 \%) is due to $\ee \rightarrow$ {\sl hadrons} and
$ \ee \ra {} \ee \tau^{+} \tau^{-}$  events.
\par
Inclusive  $\pi^0$ production is studied via the $\pi^0$ decay into two  resolved photons. A photon is
defined as an electromagnetic cluster formed by the energy deposited in at least 2 BGO crystals, with energy 
greater than 100 MeV and separated by more than 200 mrad  from the closest track. 
The number of photons detected is required to be at least 2 but not more than 20.
\par
Almost 2 $ 10 ^6$  events are selected, leading to more than 8 $ 10^7$   two 
photon  combinations. The distribution of the mass of the 
reconstructed $\gaga$ system 
shows a narrow $\pi^0$ peak.
 A fit of this peak
with a gaussian function over a Chebyshev polynomial 
background gives a resolution of 7.29\,$\pm$\,0.03 MeV.  
\par
Inclusive \ks \, production is studied via the \ks decay into $\pi^+ \, \pi^-$.
The distance, in the transverse plane,
 between the secondary vertex and the $\ee$ interaction point is required to be 
greater than 3 mm. 
The angle between the flight direction of the \ks candidate
(taken as that of the line between the interaction point and the secondary vertex
in the transverse plane)
and the total transverse momentum vector of the
two outgoing tracks must be less than 0.075 rad.
After this cut, around 5 $10^5$ events are selected. A fit of the \ks \, peak gives 
a resolution of 10\,$\pm$\,1 MeV.

\section*{Data analysis}

To evaluate the number of $\pi^0$s or $K^0_s$s, on each bin of \pt {} and $|\eta|$,
a fit is performed using a gaussian for the
signal and a Chebyshev polynomial parametrisation of the background. 
In the \pz case, to estimate the uncertainty on the background subtraction a side-band 
background subtraction is also performed.
 A three sigma region, "the signal region",
 is defined around the \pz peak. Two "background regions" are defined on each side
 of the signal region with the same width. The average number of 
$\gaga$ combinations in the background regions is subtracted from the number of
combinations in the signal region. Both methods give consistent results.
\par
The reconstruction efficiency is evaluated using the two Monte-Carlo
generators: PHOJET and PYTHIA. The kinematic limits of the cross section measurement are defined 
by imposing, at the generator level,   
$W_{\gaga} \ge 5 ~ \rm{GeV} $ for \pz and $W_{\gaga} \ge 10 ~ \rm{GeV}$ for \ks \@. 
The virtuality of the photon is limited to $Q^2 \le 1 ~ \rm{GeV} ^2 $ in PYTHIA, in
PHOJET, a limit $Q^2 \le 3 ~ \rm{GeV} ^2 $ for \pz and $Q^2 \le 8 ~ \rm{GeV} ^2 $ for \ks \, is imposed.
We have verified with PHOJET that 
the results do not depend on the  $Q^2$ cutoff. 
The number of \pz / \ks ~ reconstructed is obtained
with the same procedure used for the data. 
As the two generators reproduce equally well the experimental distributions \cite{l3tot},
a weighted average of the two is used to correct the data.

\par
The level$-$1 trigger efficiency is evaluated by comparing the response of two independent triggers: 
 the
energy trigger \cite{etrig} and the track trigger \cite{tracktrig}.
The  efficiency is   90$-$95 \%, varying with the data taking conditions during the year.
The  efficiencies of higher level triggers, calculated using prescaled events,
 vary from
85\% at low \pt {}  to  100\% at high \pt .

\par
The main uncertainty on the measured cross sections comes from the choice of Monte Carlo
generator  used to calculate the reconstruction efficiency
and from the background estimation. Half of 
the difference between the two generators and 
between the two estimations of the background are taken as systematic errors.
The uncertainties due to cut variations are negligible. 
\par Statistical and systematic errors are added in quadrature in the following. 
All the results are preliminary.

\section*{Results}

\begin{figure}[b!] 
\centerline{\epsfig{file=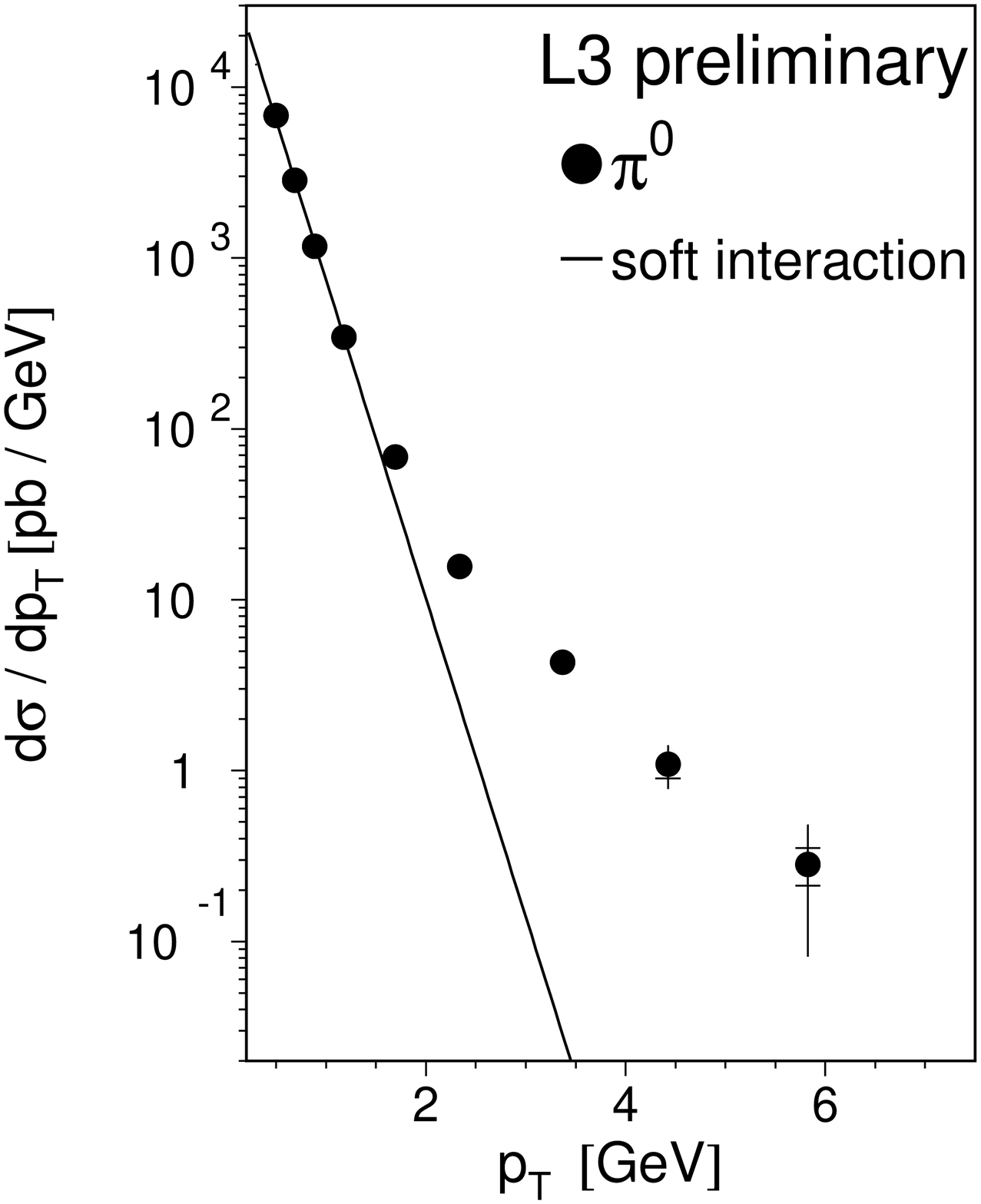,width=2.3in}
\epsfig{file=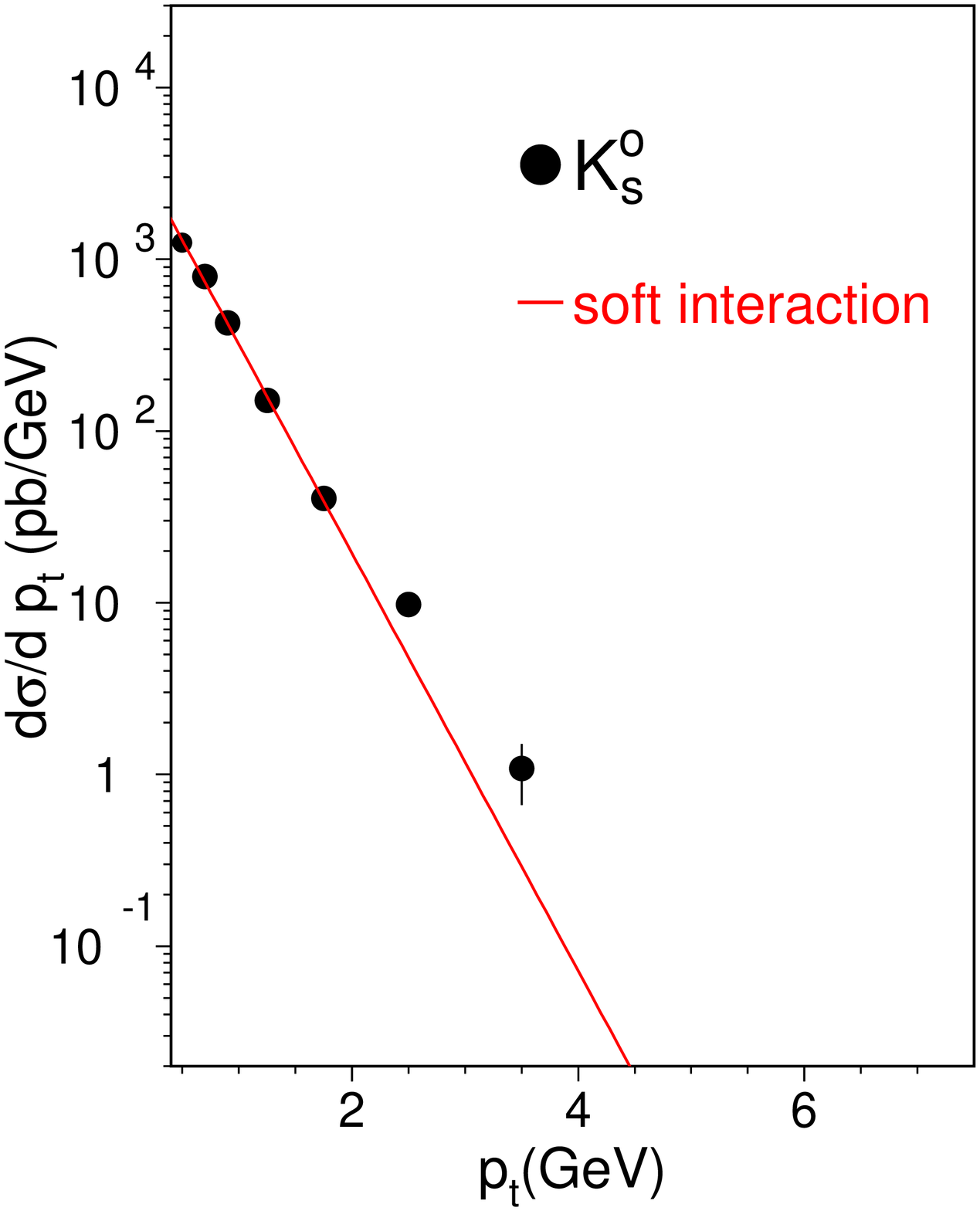,width=2.3in}}
\vspace{10pt}
\caption{The differential cross sections \dspt as a function of \pt 
for \pz and \ks \@.
The low \pt spectrum is well reproduced by an exponential fit, characteristic
of soft interactions.}
\label{fig:ptsoft}
\end{figure}

\begin{figure}[t!] 
\centerline{\epsfig{file=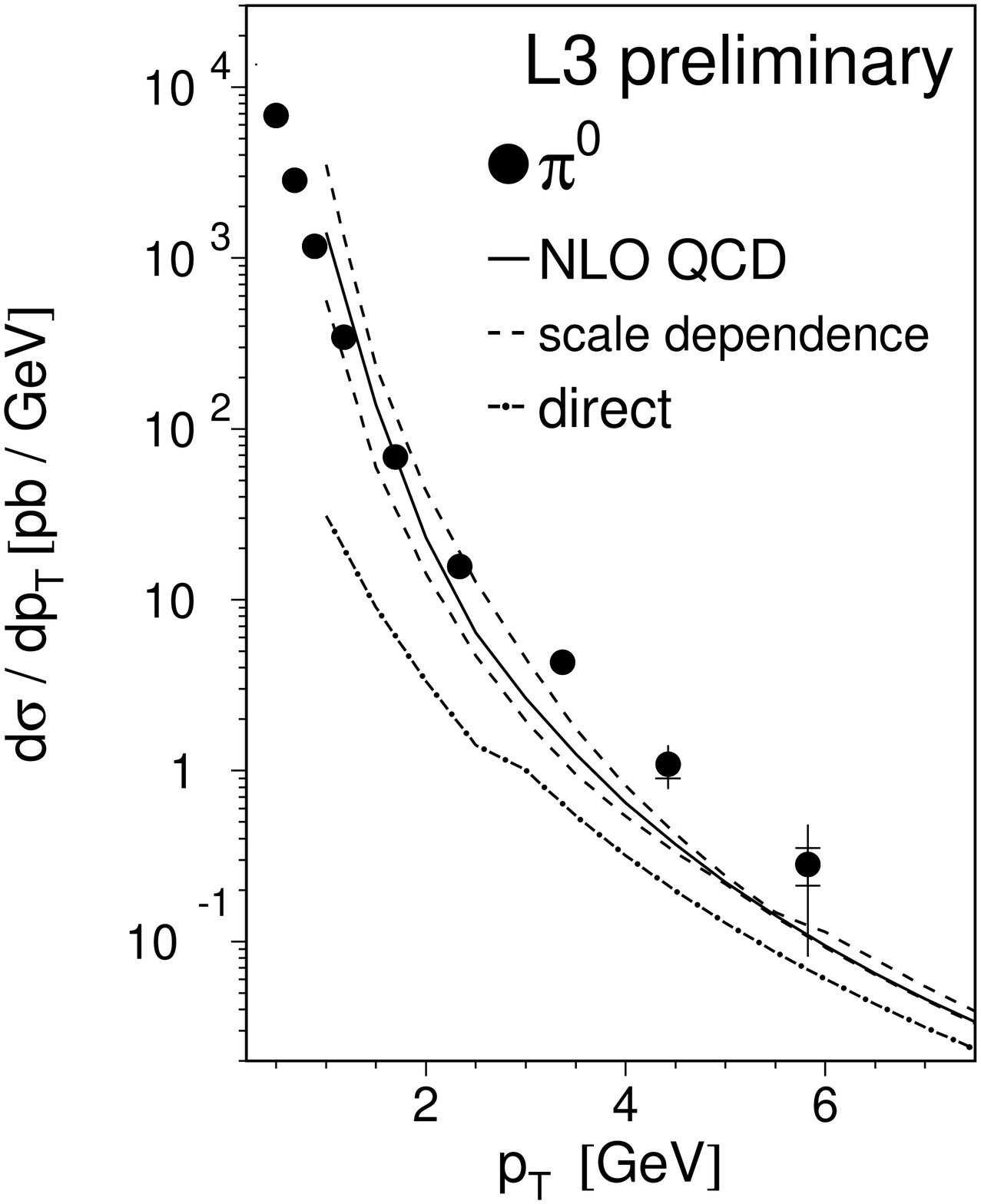,width=2.3in}
\epsfig{file=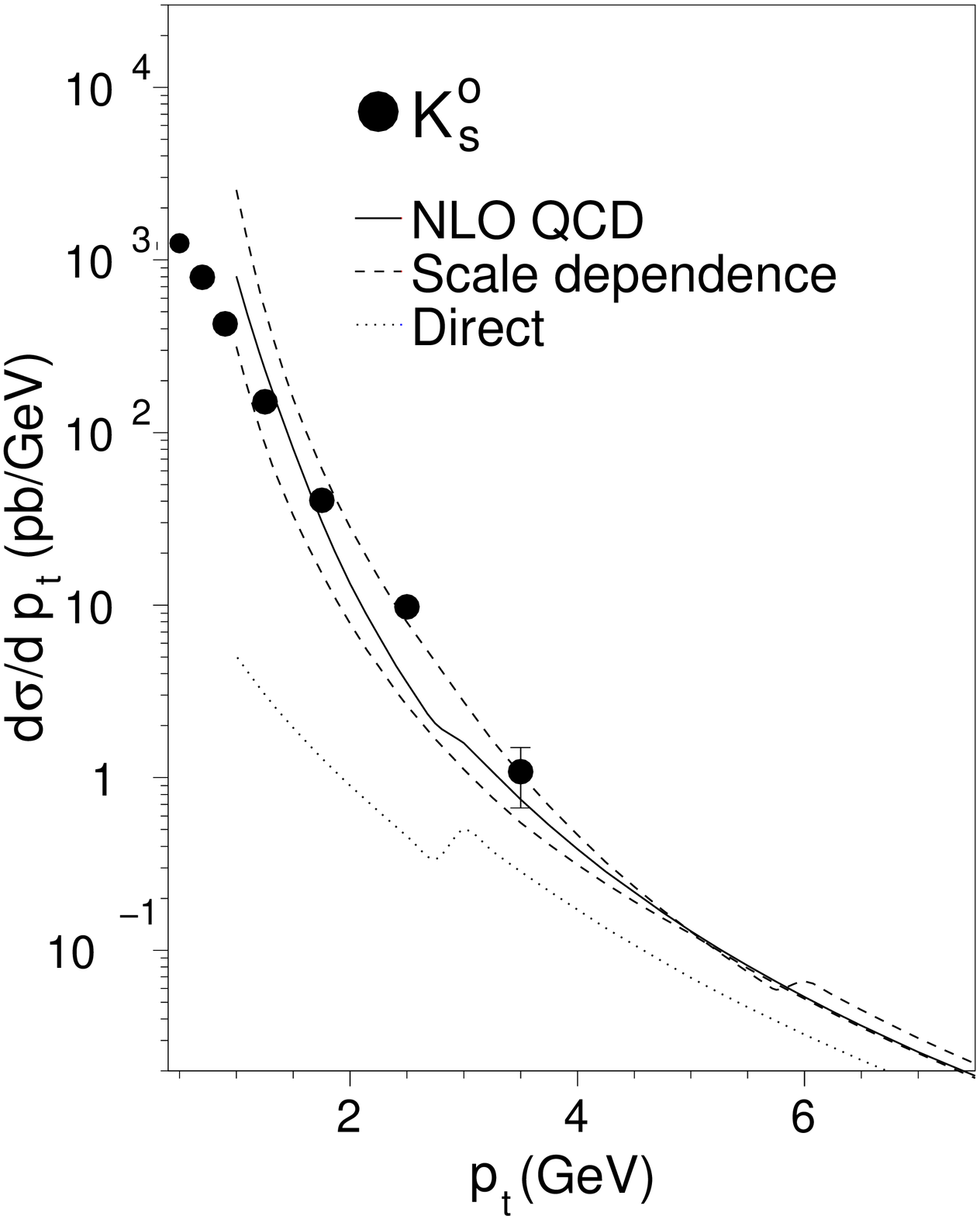,width=2.3in}}
\vspace{10pt}
\caption{Comparison of the \dspt distributions with NLO-QCD calculations.
The central values (full lines) are the calculations with $\xi=1 $ (see text),
the dashed lines refer to $\xi=0.5$ (upper lines) and to $\xi=2$ (lower lines).
The contribution of the direct process alone are indicated with dash-dotted lines.
}
\label{fig:ptQCD}
\end{figure}

\begin{figure}[b!] 
\centerline{\epsfig{file=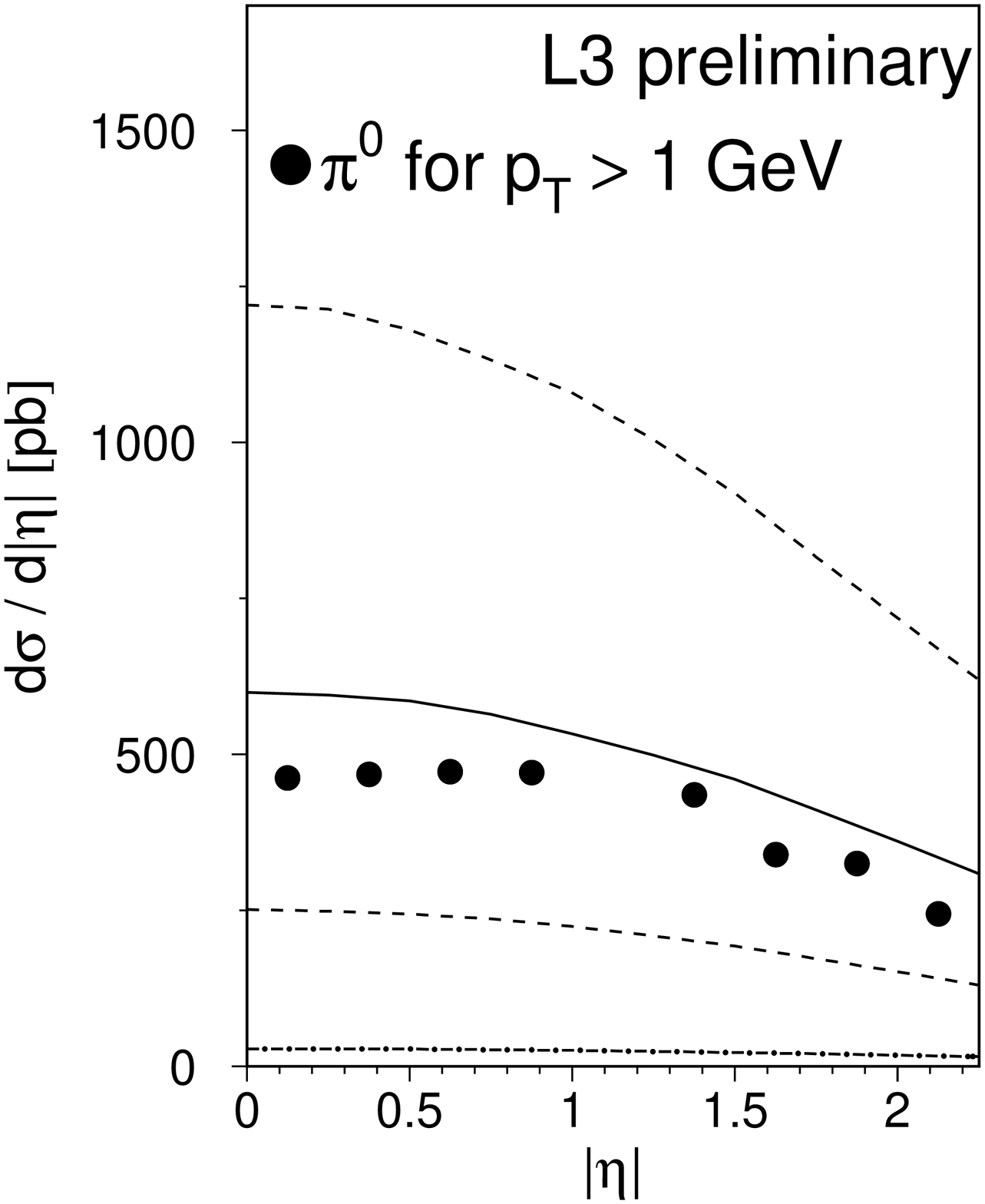,width=1.9in}
\epsfig{file=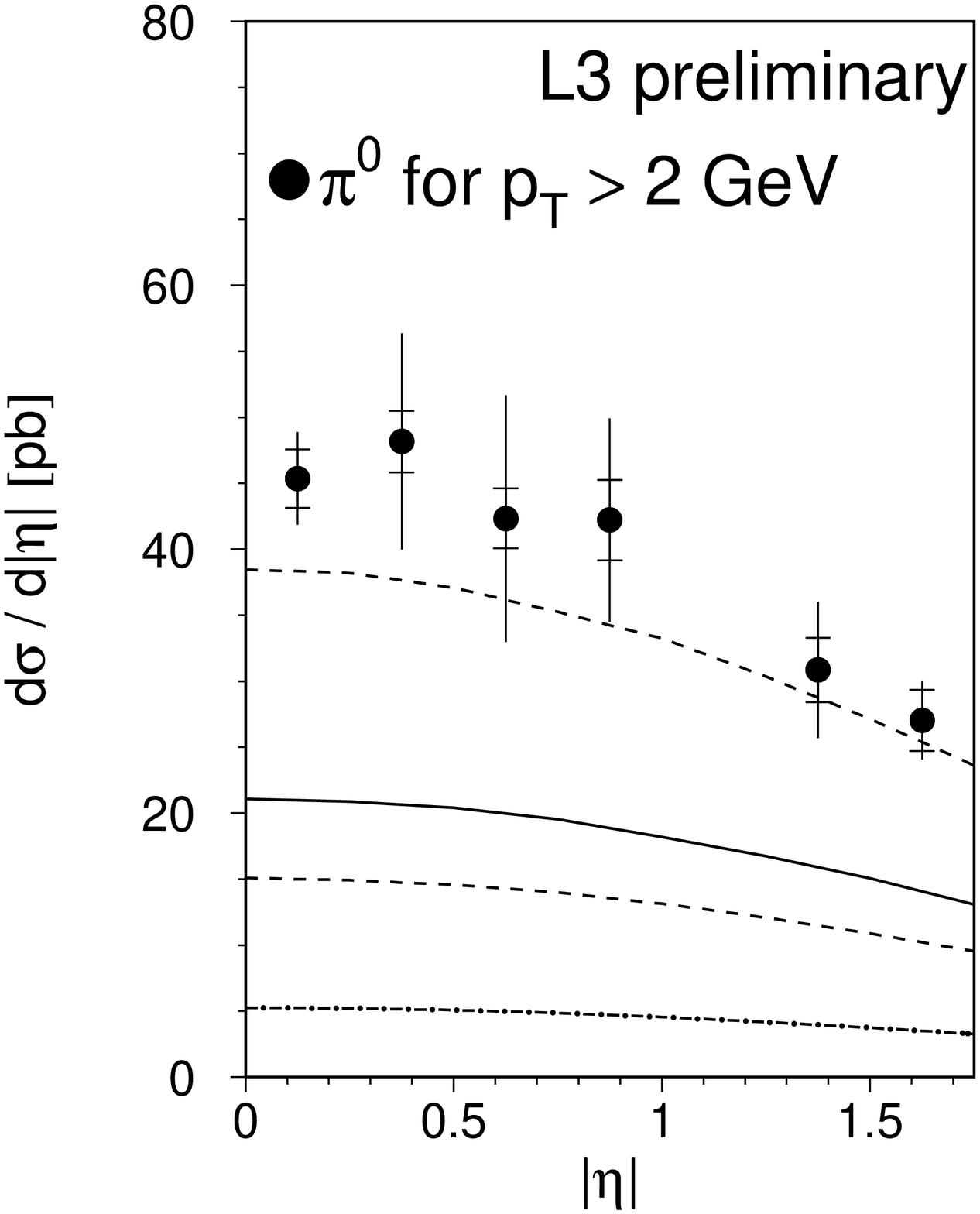,width=1.9 in}
\epsfig{file=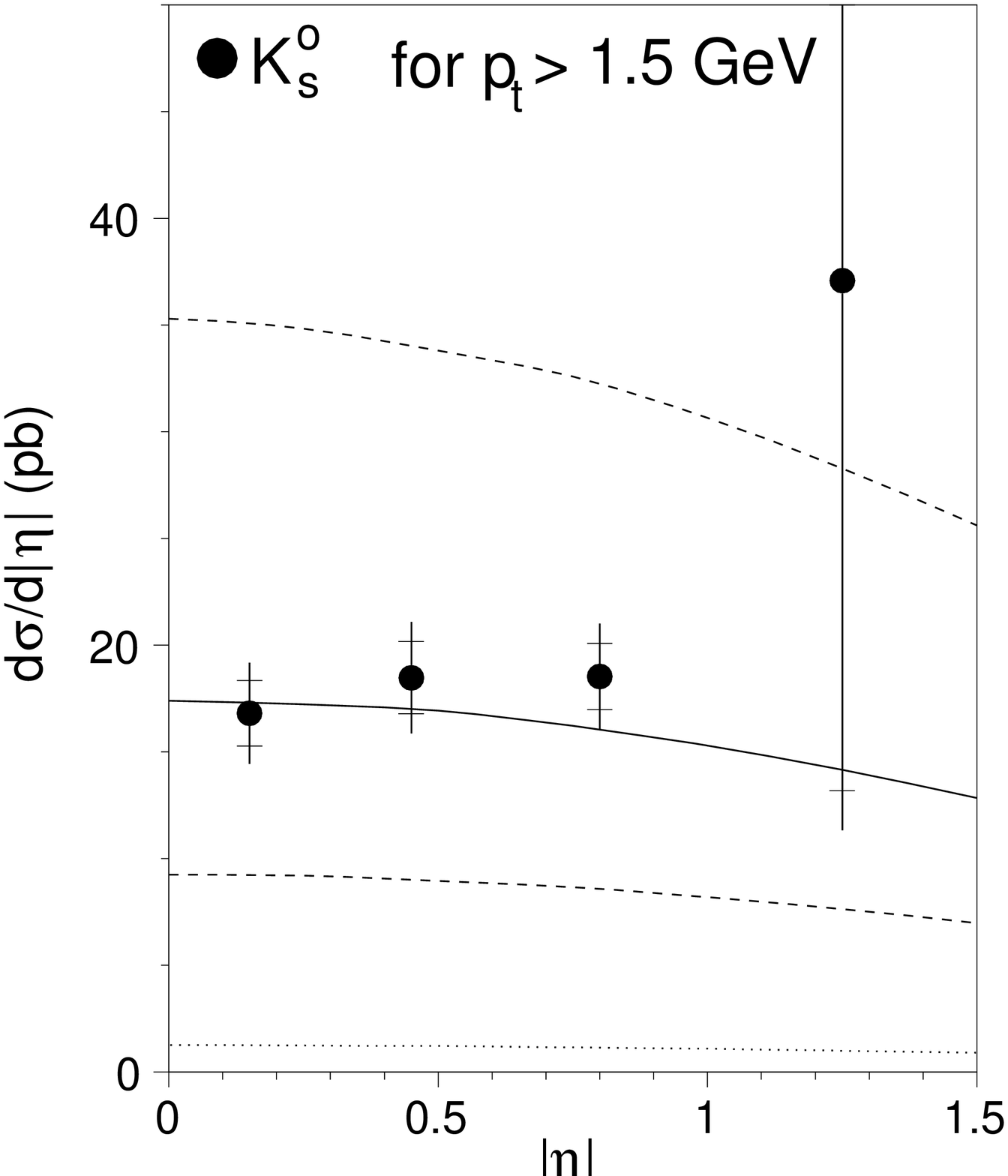,bb=0 -20 565 665, clip=true, width=1.9 in}}
\vspace{10pt}
\caption{Comparison of the \dseta distributions with NLO-QCD calculations.
The lines are as in fig.\ref{fig:ptQCD}.
}
\label{fig:eta}
\end{figure}

The differential cross sections
\dspt are shown in fig.\ref{fig:ptsoft}.
Exponential fits $e^{-p_T/<p_T>}$ for $0.2 ~ \rm{GeV} < p_T < 1.0 ~ \rm{GeV} $
reproduce well the data with $<\!p_T\!> = 233\,\pm 1~\rm{MeV}$ for \pz
and  $<\!p_T\!> = 357\,\pm 9~\rm{MeV}$ for \ks\@.
This behaviour is characteristic  of hadrons produced by soft interactions  
and the fitted values of $<\!p_T\!>$ are similar to the one obtained in hadron$-$hadron
or photon$-$hadron collisions \cite{perl}. 
\par
Due to the existence of the direct process and of hard QCD interactions, $\GG$
collisions start to exhibit a higher cross section at  \pt {} values greater than
$\sim 1.5 ~ \rm{GeV}$. 
In fig.\ref{fig:ptQCD} and \ref{fig:eta}
the \dspt and \dseta differential cross sections are compared 
to  analytical NLO QCD predictions \cite{kniehl}. For this calculation,
the flux of quasi$-$real  photons is obtained using
the Equivalent Photon Approximation, taking into account both     transverse and  longitudinal
virtual photons. 
The interacting particles can be photons or 
partons  from the $\gamma \ra q\bar{q}$ quantum fluctuation, which
evolves, via the Altarelli-Parisi equation, into quarks and gluons.
The Gordon and Storrow \cite{gordon} parton density functions are used.
All elementary $2 \ra 2$ and $2\ra 3$ processes  are considered. 
The renormalization scale, the factorisation scale 
 and   the fragmentation scale
are taken to be equal: $\mu=M=M_F=\xi p_T$.
The uncertainty in the NLO calculation is estimated by varying the value of $\xi$
from 0.5 to 2.0.

The measured cross sections are one order of magnitude 
above the direct process predictions.
In the \pz case, for \pt $> 2 ~ \rm{GeV}$, the data are also higher than the predictions 
of NLO QCD calculations.

\par
\begin{figure}[t!] 
\centerline{\epsfig{file=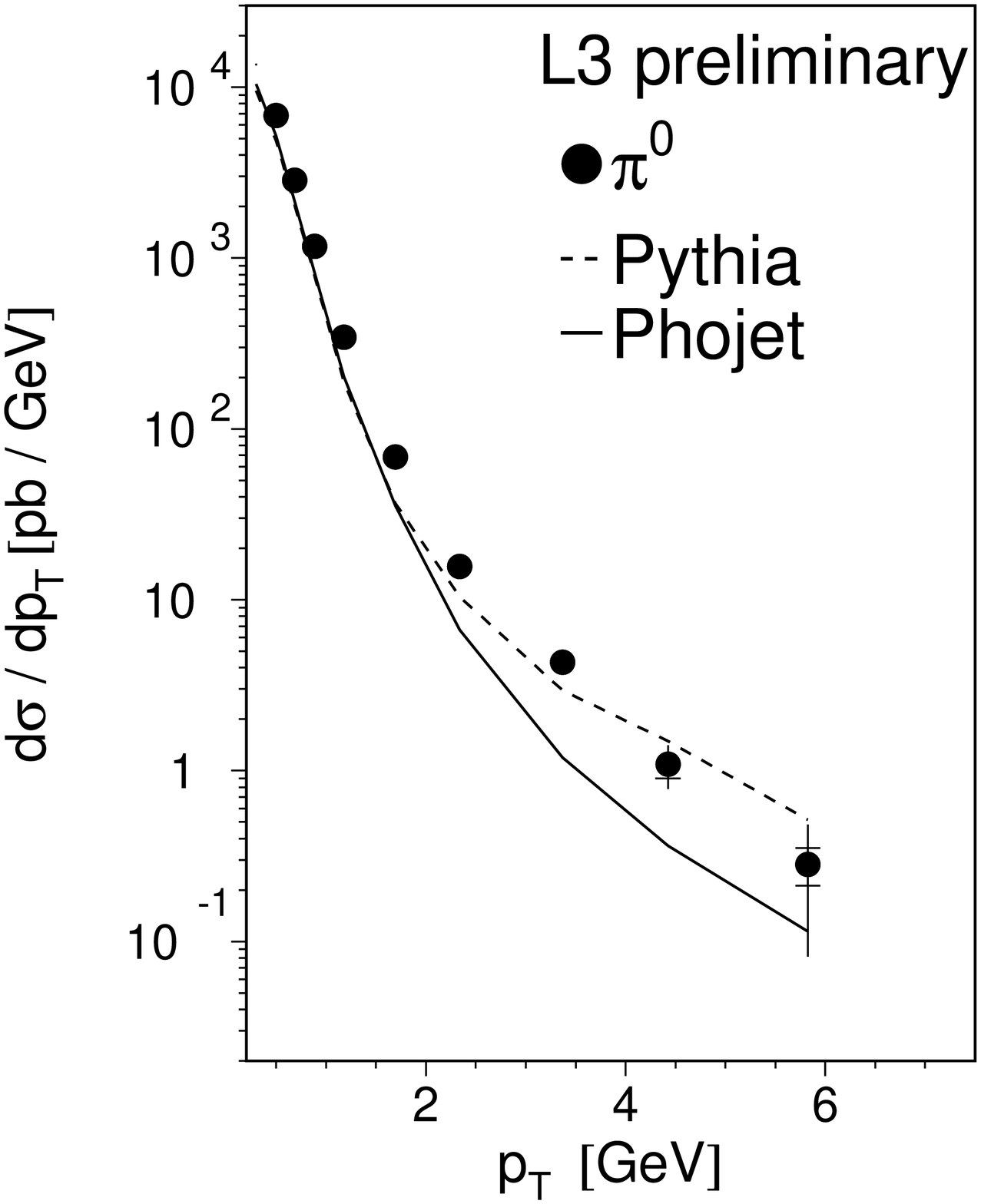,width=2.8in}
\epsfig{file=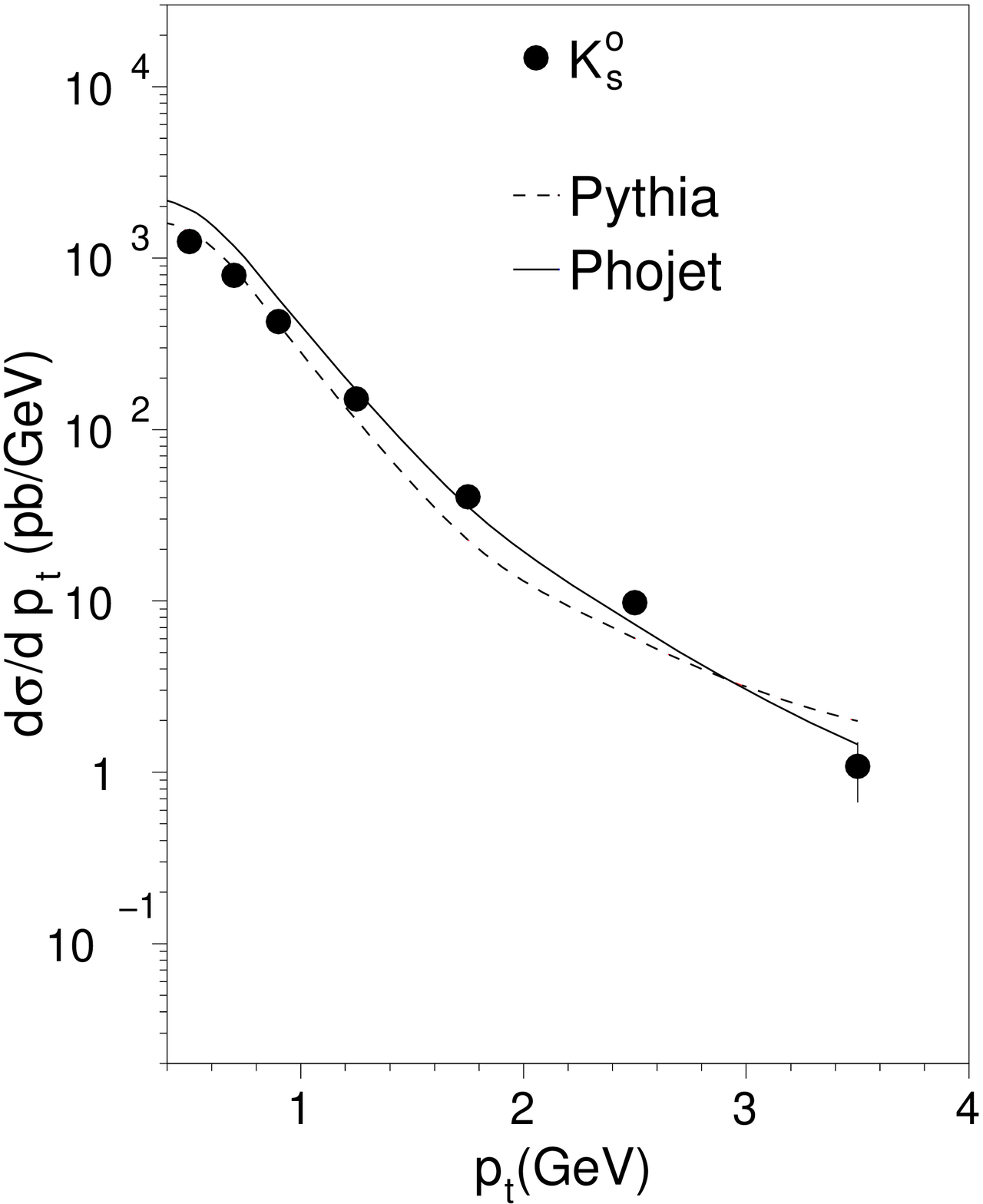,width=2.8in}}
\vspace{10pt}
\caption{Comparison of the \dspt distributions with Monte Carlo predictions.}
\label{fig:ptmc}
\end{figure}

The cross sections are also compared to Monte Carlo predictions  
 in  fig.\ref{fig:ptmc}. 
The high \pt
{} region is better reproduced by PYTHIA than by PHOJET in the \pz case.
For the \ks both Monte-Carlos reproduce quite well the data.
\par
\begin{figure}[b!] 
\centerline{\epsfig{file=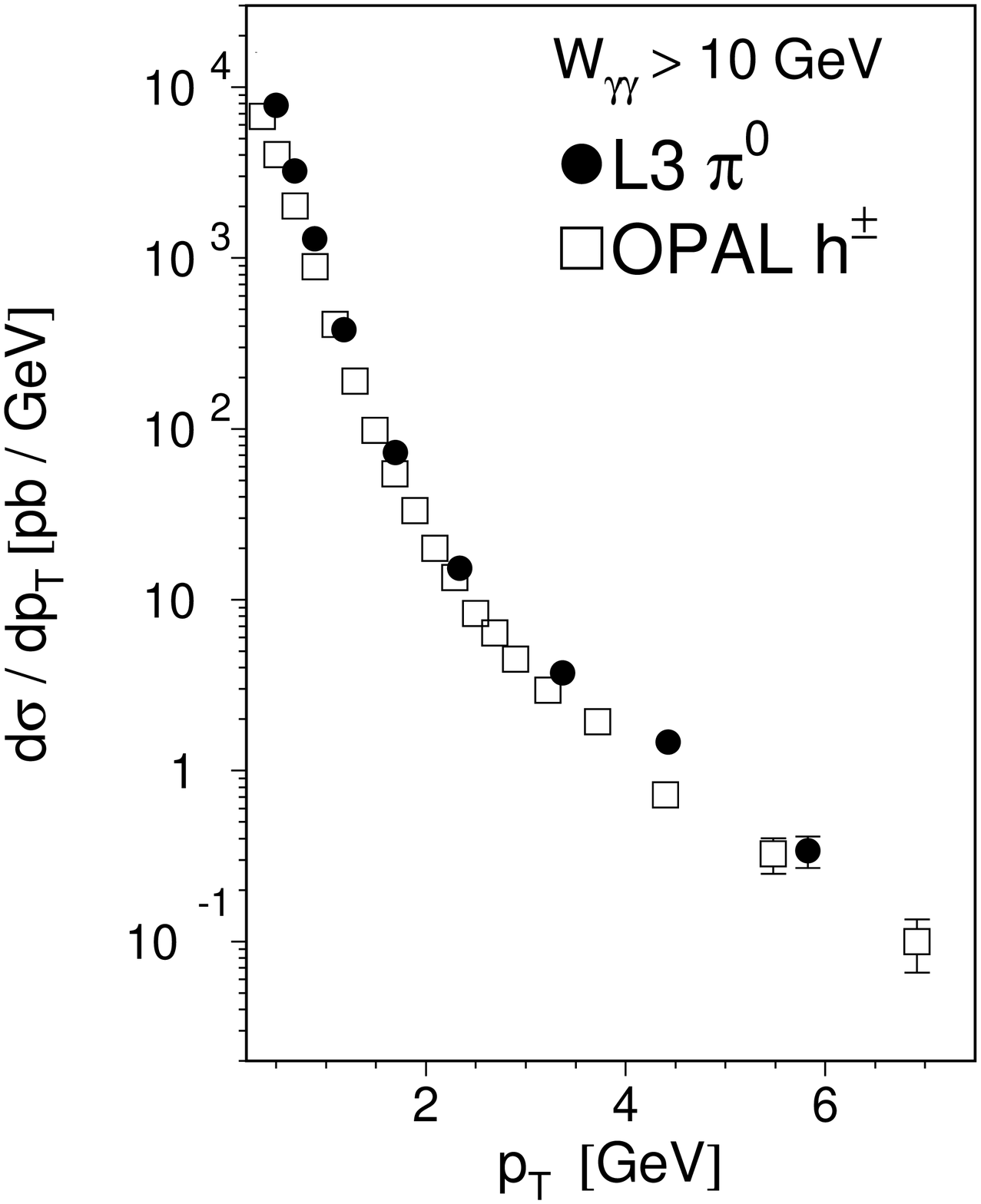,width=2.8in}
\epsfig{file=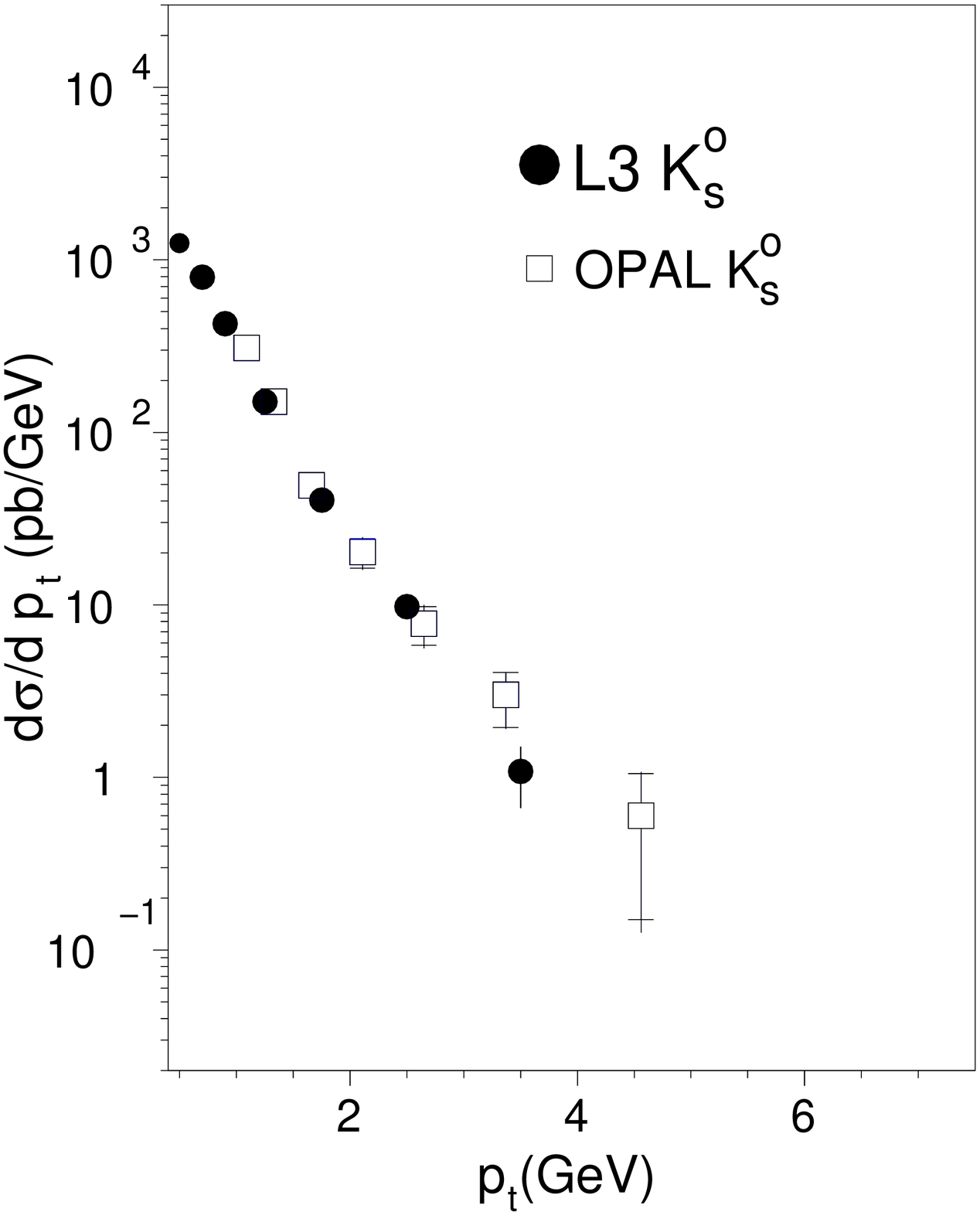,width=2.8 in}}
\vspace{10pt}
\caption{The L3 and OPAL \dspt distributions compared. The OPAL data are
scaled to the same center-of-mass energy and, in the \pz \, vs. charged
hadron case, are scaled also
to the same pseudorapidity range and fragmentation function.}
\label{fig:ptopal}
\end{figure}

The differential cross sections, calculated for $W_{\gaga} > 10 ~ \rm{GeV}$, are compared
to OPAL results \cite{opal} in fig.\ref{fig:ptopal}. OPAL measured the inclusive
production of charged hadrons (mainly $\pi^{\pm}$) 
and of \ks \, in the range $|\eta| <$ 1.5.
The experiments agree within the experimental errors.

\newpage

%
%
\clearpage
\section*{Conclusions}

This analysis is the first study of inclusive $\pi^0$  production in 
two photon collisions at LEP.
At low \pt, the differential  cross sections \dspt of \pz and \ks \,
production  follow an exponential
law similar to that previous
observed in hadron-hadron and hadron-photon collisions. At higher
\pt, the contributions of $\gaga \ra \qqbar$ and hard QCD processes are clearly 
visible. 
The agreement with previous measurement is very good, but there is only
a qualitative agreement with the predictions of LO Monte Carlo and
NLO QCD calculations.

%
%
\section*{Acknowledgments}

I would like to thank B.A. Kniehl for providing us the predictions
of NLO QCD calculations  
and L. Gordon for very useful discussions.\\
\par Thanks a lot also to A. Finch and all the Photon 2000 team for the kindness and
efficiency of the organization.\\ 
\par Many thanks, of course, to S. Braccini for providing me the \ks results.


\begin{references}
\bibitem{kniehl} Binnewies, J., Kniehl, B.A., and Kramer, G. {\it Phys. Rev.} {\bf D 53}, 6110 (1996).
\bibitem{Engel} Engel, R. , {\it Z. Phys.} {\bf C 66}, 203 (1995);\\
Engel, R.,  and Ranft, J., {\it Phys. Rev.} {\bf D 54}, 4246 (1996);\\
And Engel, R., private communication.
\bibitem{PYTHIA} Sj\"ostrand, T., {\it Comput. Phys. Commun.} {\bf 82}, 74 (1994).
\bibitem{L3} L3 Coll., Adeva, B., et al.,  {\it N.I.M.} {\bf A 289}, 35 (1990);\\
L3 Coll., Acciarri, M., et al.,  {\it N.I.M.} {\bf A 351}, 300 (1994).
\bibitem{KORALZ} Jadach, S., Ward, B. F. L., and Was, Z., 
{\it Comput. Phys. Commun.} {\bf 79}, 503 (1994).
\bibitem{KORALW} Skrzypek, M., Jadach, S., Placzek, W. and Was, Z., 
{\it Comput. Phys. Commun.} {\bf 94}, 216 (1996).
\bibitem{DIAG36} Berends, F.A., Daverfeldt, P.H., and Kleiss, R., {\it Nucl. Phys.} {\bf B 253}, 441 (1985). 
\bibitem{GEANT} Brun, R., et al., GEANT 3.15 {\it preprint CERN} DD/EE/84-1 (Revised 1987).
\bibitem{GEISHA} Fesefeldt, H., {\it RWTH Aachen report} PITHA 85/2 (1985).
\bibitem{l3tot} Kienzle, M.N., {\it these proceedings}.
\bibitem{etrig} Bizzarri, R., et al. , {\it N.I.M.} {\bf A 283}, 799 (1989).
\bibitem{tracktrig} B\'en\'e, P., et al., {\it N.I.M.} {\bf A 306}, 150 (1991);\\
Haas, D., et al., {\it N.I.M.} {\bf A 420}, 101 (1999).
\bibitem{perl} Perl, M.L., {\it High Energy Hadron Physics}, edited by J. Wiley, 1974.
\bibitem{gordon} Gordon, L.E., and Storrow, J.K., {\it Nucl. Phys.} {\bf B 489}, 405 (1997).
\bibitem{opal} OPAL Coll., Abbiendi, G., et al., {\it Eur. Phys. J.} {\bf C 6}, 253 (1999).
\end{references}
\end{document}